\documentclass{ws-ijmpa}
\usepackage[super,compress]{cite}
\usepackage{graphicx}
\usepackage{url}
\usepackage{hyperref}

\begin{document}

\title{Associated production of $J/\psi$ plus $Z(W)$ in the improved color evaporation model using the parton Reggeization approach}

\author{Alexey Chernyshev \footnote{aachernyshoff@gmail.com}  }

\address{Samara National Research University, Moskovskoe Shosse, 34, Samara, 443086, Russia}

\author{Vladimir Saleev \footnote{saleev@samsu.ru}}
\address{Samara National Research University, Moskovskoe Shosse, 34, Samara, 443086, Russia\\ and \\Joint Institute for Nuclear Research, Dubna 141980, Russia}

\maketitle


\begin{abstract}
In the article, we study associated production of prompt  $J/\psi$
mesons and $Z ( W)$ bosons in the improved color evaporation model
using the high-energy factorization as it is formulated in the
parton Reggeization approach. The last one is based on the modified
Kimber-Martin-Ryskin-Watt model for unintegrated parton distribution
functions and the effective field theory of Reggezied gluons and
quarks, suggested by L.N.~Lipatov. We predict cross section for
associated $J/\psi$ and $Z ( W)$ hadroproduction via the single and
double parton scattering mechanisms using the set of model
parameters which has been obtained early for description of single
and double prompt $J/\psi$ production at the LHC energies.  The
numerical calculations are realized using the Monte-Carlo event
generator KaTie. The calculation results are compared with the data
at the energies $\sqrt{s} = 7 \mbox{ and } 8$ TeV and we make
predictions for the energy $\sqrt{s} = 13$ TeV for $J/\psi$ plus
$Z(W)$ and $\Upsilon$ plus $Z(W)$ associated production at $\sqrt{s}
= 8$ and $13$ TeV . \vspace{0.2cm}
\end{abstract}
\vspace*{6pt}


\section{Introduction}

Associated production of prompt $J/\psi$ mesons and $Z$ or $W$
bosons is very important process for testing perturbative quantum
chromodynamics (pQCD), as well as for understanding the mechanism of
heavy quarkonium production in the high-energy collisions. The next
important issue, the experimental data for spectra and correlation
dependencies in the two particles associated production in hard
processes at the high energies, motivate us to consider not only the
conventional single parton scattering (SPS) scenario of the parton
model, but also to take into account the double parton scattering
(DPS) mechanism. There are a number of theoretical predictions for
cross sections and spectra of such processes, obtained in the
collinear parton model (CPM), which are based on the leading order
(LO)~\cite{Kniehl:2002wd} and the next-to-leading order (NLO)
approximations in the strong coupling constant
$\alpha_S$~\cite{Butenschoen:2022wld, Lansberg:2016rcx}. In
Refs.~\cite{Kniehl:2002wd,Butenschoen:2022wld}, the factorization
approach of nonrelativistic QCD (NRQCD)~\cite{NRQCD} was used to
describe the prompt $J/\psi$ plus Z-boson production. Oppositely, in
Ref.~\cite{Lansberg:2016rcx}, the one loop pQCD calculations were
performed as it is used in the color evaporation model
(CEM)~\cite{CEM1,CEM2}. In both approaches for $J/\psi$ production
mechanism, NRQCD and CEM, the results of calculations in the SPS
scenario of the NLO CPM~\cite{Butenschoen:2022wld,Lansberg:2016rcx}
are strongly underestimate the experimental data for $J/\psi$ plus
$Z$ production from the ATLAS collaboration at $\sqrt{s}= 7 \mbox{
and } 8$ TeV~\cite{ATLAS:PsiZ,ATLAS:PsiW7,ATLAS:PsiW8}. These data
as well as many others for pair heavy quarkonium production $(J/\psi
J/\psi, \Upsilon\Upsilon, \Upsilon J/\psi)$ need inclusion of large
DPS contribution in the theoretical framework.

Nowadays, the extractions of parameter $\sigma_{\rm eff}$,  based on
the DPS pocket formula, have been obtained in different experiments.
Values of $\sigma_{\rm eff}=2-25$ mb have been
derived~\cite{Chapon:2020heu}, though with large errors, with a
simple average giving $\sigma_{\rm eff} = 15$ mb. It is interesting to
study associated $J/\psi$ plus $Z(W)$ beyond the collinear
approximation of the parton model, as it was done in
\cite{Butenschoen:2022wld,Lansberg:2016rcx}, i.e. using the
high-energy factorization (HEF) or $k_T$-factorization approach
~\cite{KT1,KT2,Gribov}.

In the present study, we calculate cross section for associated
$J/\psi$ plus $Z (W)$ production in the proton-proton collisions in
the parton Reggeization approach
(PRA)~\cite{NSS2013,NKS2017,NefedovSaleev2020}, which is a version
of the $k_T$-factorization approach~\cite{KT1,KT2,Gribov}. The PRA
is accurate both in the collinear limit, which drives the
transverse-momentum-dependent (TMD) factorization~\cite{CPM} and in
the high-energy limit, which is important for
Balitsky-Fadin-Kuraev-Lipatov~(BFKL)~\cite{BFKL1,BFKL2,BFKL3,BFKL4}
resummation of $\ln (\sqrt{s} / \mu)$-enhanced effects. In the PRA,
we have studied successfully a heavy quarkonium production in the
proton-(anti) proton collisions at the Tevatron and the LHC using
NRQCD approach, see Refs.~\cite{VKS2005A,VKS2005B,NSS2012,NS2016}.
The pair production of prompt $J/\psi$ was studied in the PRA using
NRQCD approach in the Ref.~\cite{He:2019qqr} and using the improved
color evaporation model (ICEM)~\cite{ICEM_Ma_Vogt} in the
Ref.~\cite{ChernyshevSaleev2Psi2022}. In our previous study of
single and double $J/\psi$ or $\Upsilon$ production, as well as
$J/\psi  \Upsilon$ pair production, in the PRA using the ICEM
~\cite{ChernyshevSaleev2Psi2022, ChernyshevSaleev2Ups2022}, we
considered both, the SPS and the DPS production mechanism, and we
have found numerical values for parameters of the ICEM
$({\cal F}^{\psi}, {\cal F}^{\Upsilon}$) at the energy of LHC and the
universal DPS parameter $\sigma_{\rm eff}$. These parameters are used
during the calculations presented below.

The paper has the following structure. In Section~\ref{sec:models},
the PRA formalism  and  the ICEM are shortly reviewed. The main
formulas of the SPS and DPS scenarios are presented as well as
details of the numerical calculations which are based on the
Monte-Carlo parton-level event generator KaTie~\cite{katie}. In
Section~\ref{sec:results}, we compare our predictions and the
experimental data for the associated production of prompt $J/\psi$
mesons and $Z(W)$ bosons at the energies $7$ and $8$ TeV . We also
present predictions for the energy $\sqrt{s} = 13$ TeV, both for the
$J/\psi$ plus $Z(W)$ production and for $\Upsilon$ plus $Z(W)$
associated production. Our conclusions are summarized in
Section~\ref{sec:conclusions}.

\section{Models and numerical methods}
\label{sec:models}
\subsection{The parton Reggeization approach}
The PRA is based on the HEF or $k_T$-factorization justified in the
leading logarithmic approximation of the QCD at high
energies~\cite{KT1,KT2,Gribov}. Dependent on transverse momentum,
parton distribution functions (PDFs) of Reggeized quarks and gluons,
which are considered in the PRA, are calculated in the Kimber,
Martin, Ryskin and Watt (KMRW) model~\cite{KMR,WMR}, but with
sufficient modifications~\cite{NefedovSaleev2020} that will be
described below.

Reggeized parton amplitudes in the PRA are constructed according to
the Feynman rules of the L.N.~Lipatov Effective Field Theory (EFT)
of Reggeized gluons and quarks~\cite{Lipatov95, LipatovVyazovsky}.
That is guarantee of their gauge invariance. A detailed
description of the PRA can be found in
Refs.~\cite{NSS2013,NKS2017,NefedovSaleev2020}, inclusion of
corrections from the emission of additional partons to the leading
PRA approximation was studied in the Refs.~\cite{NKS2017, NS2019},
the development of the PRA with loop corrections was considered in
the Refs.~\cite{PRANLO1,PRANLO2,PRANLO3}.

In the PRA, the cross-section of the process
$p + p \to {\cal H} + X$ is related to the cross-section of the
parton subprocess $i + j \to {\cal H}+X$ by the factorization formula
\begin{eqnarray}
  d\sigma & = & \sum_{i, \bar{j}}
    \int\limits_{0}^{1} \frac{dx_1}{x_1} \int \frac{d^2{\bf q}_{T1}}{\pi}
{\Phi}_i(x_1,t_1,\mu^2)
    \int\limits_{0}^{1} \frac{dx_2}{x_2} \int \frac{d^2{\bf q}_{T2}}{\pi}
{\Phi}_{j}(x_2,t_2,\mu^2)\cdot d\hat{\sigma}_{\rm PRA},
\label{eqI:kT_fact}
\end{eqnarray}
where $t_{1,2} = - {\bf{q}}_{T 1,2}^2$, the cross-section of the
subprocess with Reggeized partons $d\hat{\sigma}_{\mathrm{PRA}}$ is
expressed in terms of squared Reggeized amplitudes
$\overline{|{\mathcal{A}}_{\mathrm{PRA}}|^2}$ in the standard way.

Unintegrated PDFs (unPDFs) in the modified KMRW model are calculated
by the formula~\cite{NefedovSaleev2020}
\begin{equation}
\Phi_i(x,t,\mu) = \frac{\alpha_s(\mu)}{2\pi}
\frac{T_i(t,\mu^2,x)}{t}
\sum\limits_{j = q,\bar{q},g}\int\limits_{x}^{1} dz\ P_{ij}(z) {F}_j
\left(
\frac{x}{z}, t \right) \theta\left( \Delta(t,\mu)-z
\right),\label{uPDF}
\end{equation}
where $F_i(x,\mu_F^2) = x f_j(x,\mu_F^2)$. Here and below, we put
factorization and renormalization scales are equal, $\mu_F = \mu_R =
\mu$, and $\Delta(t,\mu^2)=\sqrt{t}/(\sqrt{\mu^2}+\sqrt{t})$ is the
KMRW-cutoff function~\cite{KMR}. To resolve collinear divergence
problem, we require that the modified unPDF ${\Phi}_i(x,t,\mu)$
should be satisfied exact normalization condition:
\begin{equation}
\int\limits_0^{\mu^2} dt \Phi_i(x,t,\mu^2) =
{F}_i(x,\mu^2),\label{eq:norm}
\end{equation}
or
\begin{equation}
\Phi_i(x,t,\mu^2) = \frac{d}{dt}\left[ T_i(t,\mu^2,x){F}_i(x,t)
\right],\label{eq:sudakov}
\end{equation}
where $T_i(t,\mu^2,x)$ is the Sudakov form--factor, $T_i(t =
0,\mu^2,x) = 0$ and $T_i(t = \mu^2,\mu^2,x) = 1$. The explicit form
of the Sudakov form factor in the ({\ref{eq:sudakov}) was first
obtained in~\cite{NefedovSaleev2020}:
\begin{equation}
T_i(t,\mu^2,x) = \exp\left[-\int\limits_t^{\mu^2} \frac{dt'}{t'}
\frac{\alpha_s(t')}{2\pi} \left( \tau_i(t',\mu^2) + \Delta\tau_i
(t',\mu^2,x) \right) \right],\label{eq:sud}
\end{equation}
where
\begin{eqnarray*}
\tau_i(t,\mu^2) & = & \sum\limits_j \int\limits_0^1 dz\ zP_{ji}(z)\theta
(\Delta(t,\mu^2)-z), \label{eq:tau} \\
\Delta\tau_i(t,\mu^2,x) & = & \sum\limits_j \int\limits_0^1 dz\
\theta(z - \Delta(t,\mu^2)) \left[ zP_{ji}(z) -
\frac{{F}_j\left(\frac{x}{z},t \right)}{{F}_i(x,t)} P_{ij}(z)
\theta(z - x) \right].
\end{eqnarray*}
In contrast to the KMRW model, the Sudakov form factor
(\ref{eq:sud}) depends on $x$, which is necessary to preserve the
exact normalization (\ref{eq:norm}) for any $x$ and $\mu$. The gauge
invariance of amplitudes with Reggeized partons  in the PRA
guaranteed allows you to study any processes described non-Abelian
QCD structures.

\subsection{Improved Color Evaporation Model} \label{sec:icem}
First time, the CEM was proposed many years ago in
Refs.~\cite{CEM1,CEM2}. Later, the CEM was improved by Ma and
Vogt~\cite{ICEM_Ma_Vogt} and is now used to describe the spectra
and polarizations of $J/\psi$-mesons in the collinear parton model
(CPM)~\cite{ICEM2017,ICEM2021} and in the $k_T$-factorization
approach~\cite{ICEM_KT1,ICEM_KT2}. The current status of the ICEM is
presented in the Ref.~\cite{ICEM_Ma_Vogt}.

In the combine approach of the PRA and ICEM, the leading order (LO)
parton subprocesses of the associated $J/\psi$ and $Z$ boson production, in
the strong interaction constant $\alpha_S$, via the SPS scenario are the
following:
\begin{align}
R + R & \to c + \bar{c} + Z, \label{RRccZ} \\
Q_q + \bar Q_q & \to c +\bar c + Z,  \label{QQccZ}
\end{align}
where $R$ is a Reggeized gluon, $Q_q(\bar{Q}_q)$ is a Reggeized
quark (antiquark) and $q = u, d, s, c, b$. The $W^\pm$ bosons are
produced in the LO approximation via the quark-antiquark scattering
\begin{eqnarray}
&& Q_u + \bar Q_d \to c + \bar c + W^+, \\ \label{udW}
&& Q_d + \bar Q_u \to c + \bar c + W^-, \\ \label{duW}
&& Q_c + \bar Q_s \to c + \bar c + W^+, \\ \label{csW}
&& Q_s + \bar Q_c \to c + \bar c + W^- \label{scW},
\end{eqnarray}
where we neglect small contributions from  subprocesses, which are
suppressed by the factor $\sin^2 (\theta_W) \simeq 0.04$.

In the ICEM, the cross section for the production of prompt $J/\psi$
mesons is related to the cross section for the production of
$c \bar{c}$-pairs in the SPS scenario as follows:
\begin{eqnarray}
\sigma^{\rm SPS}(p + p \to J/\psi + Z(W) + X)& =& \mathcal{F}^\psi
\times \\ \nonumber && \times \int_{m_\psi}^{2m_D} \frac{d \sigma(p
+ p \to c + \bar c + Z(W) + X)}{dM} d M, \label{eq:ICEM1}
\end{eqnarray}
where $M$ is the invariant mass of the $c \bar{c}$ pair with
4-momentum $p_{c \bar{c}}^\mu = p_c^\mu + p_{\bar{c}}^\mu$,
$m_\psi$ is the mass of the $J/\psi$ meson and $m_D$ is the mass of
the lightest $D$ meson. To take into account the kinematic effect
associated with the difference between the masses of the
intermediate state and the final charmonium, the 4-momentum of
$c \bar{c}$ pair and $J/\psi$ meson are related by
$p_\psi^\mu = (m_\psi / M) \, p_{c \bar{c}}^\mu$. The universal parameter
$\mathcal{F}^\psi$ is considered as a probability of transformation
of the $c\bar{c}$ pair with invariant mass $m_{\psi} < M < 2 m_D$
into the prompt $J/\psi$ meson. During the numerical calculations
we use parameters ${\cal F^\psi}$ and ${\cal F}^\Upsilon$, which
have been found in the ICEM fit for single prompt $J/\psi$ and
$\Upsilon$ production cross sections at the energies $\sqrt{s}=7 - 13$
TeV. These parameters can be founded in the Table I of Ref.
\cite{ChernyshevSaleev2Psi2022} for the prompt $J/\psi$ and in Fig.
3 of Ref.~\cite{ChernyshevSaleev2Ups2022} for the prompt $\Upsilon$
production. For the readers convenience, we summarize these results
in the Table.

In the DPS approach~\cite{DPS}, the cross section for the associated
production of $J/\psi$ plus $Z/W$ is written in terms of the cross
sections for the production of a prompt $J/\psi$ meson and the
$Z / W$ bosons in two independent subprocesses
\begin{eqnarray}
\sigma^{\mathrm{DPS}}(p + p \to J/\psi + Z (W) + X)= \frac{1}{
\sigma_{\mathrm{eff}}} && \times \sigma^{\mathrm{SPS}}(p + p \to
J/\psi + X_1) \\ \nonumber && \times  \sigma^{\mathrm{SPS}}(p + p
\to Z(W) + X_2),
\end{eqnarray}
where the parameter $\sigma_{\mathrm{eff}}$ controls the
contribution of the DPS mechanism and it is taken as it was obtained
in the heavy quarkonium pair production in
Refs.~\cite{ChernyshevSaleev2Psi2022,ChernyshevSaleev2Ups2022},
$\sigma_{\mathrm{eff}} = 11.0\pm 0.2$ mb. Thus, in the presented
calculations we don't use any free parameters to predict the cross
sections for associated heavy quarkonium plus $Z( W)$ production.

To calculate the single prompt $J/\psi$ cross section in the ICEM,
we consider following subprocesses:
\begin{eqnarray}
R + R & \to c + \bar c, \\
Q_q + \bar Q_q & \to c + \bar c, \end{eqnarray}
To calculate inclusive cross section
$\sigma(p + p \to Z / W + X)$, we include the following leading parton
subprocesses in our analysis:
\begin{align}
Q_q + \bar Q_q & \to Z, \label{qqZ} \\
Q_u + \bar Q_d & \to W^+, \label{Wq1} \\
Q_d + \bar Q_u & \to W^-, \label{Wq2} \\
Q_c + \bar Q_s & \to W^+, \label{Wq3} \\
Q_s + \bar Q_c & \to W^-. \label{Wq4}
\end{align}
Here we should cite on our previous calculations of the inclusive
$Z$ boson production in the PRA~\cite{NefedovSaleev2020}. It was
shown that in the calculation including only LO subprocess
(\ref{qqZ}), we describe well the normalized transverse momentum
spectrum of $Z$ mesons, but to described the total cross section we
have to multiply LO result by the $K$-factor, which is about $K
\simeq 1.6$. During the presented above calculations we use
$K$-factors to calculate inclusive $Z ( W)$ boson production cross
section as it is written in the Table \ref{table:1}.

    \begin{table}[h]
    \centering

\begin{tabular}{c c c c c c}

\hline\hline  $\sqrt{s}$, TeV & Cross section & Data [nb] & LO PRA
[nb]
& $K^{\rm exp}_{\rm theor}$ \\
\hline

$8$ & $\sigma(Z)$ & $33.14$ & $20.80$ & $1.59$ \\

 $13$ &$\sigma(Z)$ & $40.31$ & $28.03$ & $1.43$ \\

$7$ & $\sigma(W^{\pm})$ & $98.71$ & $66.59$ & $1.51$ \\

 $8$ & $\sigma(W^{\pm})$ & $112.43$ & $75.97$ & $1.48$ \\

 $13$& $\sigma(W^\pm)$ & $152.07$ & $103.34$ & $1.51$ \\

\hline\hline

\end{tabular}

    \caption{Comparison of experimental and theoretical cross sections for
    inclusive $Z$ and $W^{\pm}$ productions. The data are from
    CMS Collaboaration~\cite{CMSinclZW8} and
    ATLAS Collaboration~\cite{ATLASinclW7, ATLASinclZ13, ATLASinclW13}.}
    \label{table:1}
    \end{table}

\subsection{Numerical methods} \label{sec:methods}
As demonstrated in the Ref.~\cite{ChernyshevSaleev2Psi2022},
we may apply fully numerical method of the calculation in the PRA
using the Monte Carlo (MC) parton level event generator
KaTie~\cite{katie}. The approach to obtaining gauge invariant
amplitudes with off-shell initial state partons in scattering at
high energies was proposed in the Ref.~\cite{hameren1,hameren2}. The
method is based on the use of spinor amplitudes formalism and
recurrence relations of the Britto-Cachazo-Feng-Witten (BCFW) type.
In Ref.~\cite{katie}, the Monte Carlo (MC) parton level event
generator KaTie for processes at high energies with nonzero
transverse momenta and virtualities was developed. This
formalism~\cite{katie,hameren1,hameren2} for numerical amplitude
generation is equivalent to amplitudes built according to Feynman
rules of the Lipatov EFT at the level of tree
diagrams~\cite{NSS2013,NKS2017,kutak}. The accuracy of numerical
calculations using KaTie for total proton-proton cross sections is
equal to 0.1\%.

Concerning the calculation of the $Z$ boson production using KaTie
we should make a special explanation. In the
experiment~\cite{CMSinclZW8}, it is studied the lepton pair
production, $p + p \to \ell + \bar \ell + X \; (\ell = e, \mu)$
processes, in the range of two-lepton invariant mass near the Z boson peak
$|M_{\ell \bar \ell} - m_Z| < 10$ GeV, where contribution of the subprocess
$q + \bar q \to Z \to \ell + \bar \ell$ dominates over the photon-exchange subprocess $q + \bar q \to \gamma^* \to \ell + \bar \ell$.
Such a way, instead of the subprocesses (\ref{RRccZ}) and (\ref{QQccZ}), we calculate the following ones
\begin{align}
R + R & \to c + \bar{c} + \ell + \bar \ell, \label{RRccMU} \\
Q_q + \bar Q_q & \to c + \bar c + \ell + \bar \ell  \label{QQccMU}.
\end{align}
Respectively, instead of subprocess (\ref{qqZ}), which is used in
the DPS approach, we calculate next one
\begin{equation}
Q_q+\bar Q_q \to \ell + \bar \ell \label{qqMU}\\
\end{equation}
In case of $W^\pm$ production, we calculate cross section of
$\mu + \nu_\mu$ or $\bar \mu + \bar\nu_\mu$ production in the
subprocesses
\begin{eqnarray}
&& Q_u + \bar Q_d \to \mu + \nu_\mu + c + \bar c, \\ \label{udNU}
&& Q_d + \bar Q_u \to \bar \mu + \bar \nu_\mu + c + \bar c, \\ \label{duNU}
&& Q_c + \bar Q_s \to \mu + \nu_\mu + c + \bar c ,\\ \label{csNU}
&& Q_s + \bar Q_c \to \bar \mu + \bar \nu_\mu + c + \bar c
\label{scNU}.
\end{eqnarray}
We separate events of $W^\pm$ production as it is described in the
Refs.~\cite{ATLAS:PsiW7,ATLAS:PsiW8}. The phase-space cuts of
measurements for associated $J/\psi$ plus $Z/W$ production are
collected in the Table~\ref{table:2}.

\def\arraystretch{1.2}

\begin{table}[h]

\centering
\begin{tabular}{l c l}

\hline\hline Collaboration & Rapidity &
Transverse Momentum \\
\hline

\multicolumn{2}{c}
{$J/\psi + Z \left( \to \ell \bar\ell \right)$} \\
\hline

ATLAS, $\sqrt{s} = 8$ TeV: &
 $|y^{\psi}| < 2.1$ & $p_T^{\psi} \in [8.5, 100]$ GeV \\
 \hline
 \multicolumn{3}{c}{$Z$ cuts} \\
 \hline
 \multicolumn{1}{c}{$|M_{\ell_1 \ell_2} - m_Z| < 10$ GeV}
 & $|\eta^{\ell_1}| < 2.5$ & $p_T^{\ell_1} > 25$ GeV \\
 & $|\eta^{\ell_2}| < 2.5$ & $p_T^{\ell_2} > 15$ GeV \\
 & & $p_T^{\ell_1} > p_T^{\ell_2}$ \\

\hline\hline

\multicolumn{2}{c}
{$J/\psi + W^{\pm} \left( \to \ell \nu_{\ell} \right)$} \\
\hline

ATLAS, $\sqrt{s} = 7$ TeV:
&
 $|y^{\psi}| < 2.1$ & $p_T^{\psi} \in [8.5, 30]$ GeV \\
 \hline
 \multicolumn{3}{c}{$W^{\pm}$ cuts} \\
 \hline
 \multicolumn{1}{c}{$M_T^{W} > 40$ GeV} & $|\eta^{\ell}| < 2.4$
 & $p_T^{\ell} > 25$ GeV \\
 & & $p_T^{\nu} > 20$ GeV \\

\hline\hline

\multicolumn{2}{c}
{$J/\psi + W^{\pm} \left( \to \ell \nu_{\ell} \right)$} \\
\hline

ATLAS, $\sqrt{s} = 8$ TeV:
&
 $|y^{\psi}| < 2.1$ & $p_T^{\psi} \in [8.5, 150]$ GeV\\
 \hline
 \multicolumn{3}{c}{$W^{\pm}$ cuts} \\
 \hline
 \multicolumn{1}{c}{$M_T^{W} > 40$ GeV} & $|\eta^{\ell}| < 2.4$
 & $p_T^{\ell} > 25$ GeV \\
 & & $p_T^{\nu} > 20$ GeV \\
\hline\hline

\end{tabular}

    \caption{Phase-space cuts of
    measurements~\cite{ATLAS:PsiZ,ATLAS:PsiW7,ATLAS:PsiW8}
    of associated production of prompt $J/\psi$ mesons and $Z ( W)$ bosons,
    $m_Z = 91.1876$ GeV and $m_W = 80.379$ is used. In case of $Z$ decays
    $\ell = e, \mu$, $\ell_1$ is the leading lepton and $\ell_2$ is the
    subleading lepton $\left( p^{\ell_1}_T > p^{\ell_2}_T \right)$,
    and in case of $W^{\pm}$ decays $\ell = \mu, \bar\mu$,
    $M_T^W$ is called as the transverse energy of the $W$ boson and
    defined as
    $M^W_T = \sqrt{2 \, p_T^{\mu} \, p_T^{\nu} \,
    (1 - \cos\Delta\phi_{\mu\nu})}$. }

    \label{table:2}
    \end{table}

\section{Results}
\label{sec:results} Now we are in position to compare our
theoretical predictions obtained in the ICEM using the PRA within
the SPS and the DPS contributions  with the
data~\cite{ATLAS:PsiZ,ATLAS:PsiW7,ATLAS:PsiW8} . First of all, we
describe phase-space definition of the measured fiducial production
cross-section following the geometrical acceptance of the ATLAS
detector~\cite{ATLAS:PsiZ} at $\sqrt{s} = 8$ TeV. The Z boson
selection via two-lepton decays $Z \to \ell \bar \ell \;
(\ell=e,\mu)$ includes following criteria: pseudorapidity of each
lepton $|\eta^{\ell}| < 2.5$, leading lepton has $p_{T}^{\ell_1} >
25$ GeV and subleading lepton has $p_{T}^{\ell_2} > 15$ GeV,
$|M_{\ell \bar \ell} - m_Z| < 10$ GeV, where $M_{\ell \bar \ell}$ is
the invariant mass of the lepton pair. Associated with Z boson
prompt $J/\psi$ meson satisfies the following conditions:
$|y^{\psi}| < 2.1$ and $8.5 < p_T^{\psi} < 100$ GeV. Secondly, in
case of $J/\psi + W^{\pm}$ production, $W^{\pm}$ is reconstructed in
the lepton-neutrino decays $W^{\pm} \to \ell \nu_{\ell} \; (\ell =
\mu, \bar\mu)$. To separate events from $W^{\pm}$ decays criteria
are as follows: leptons have $|\eta^{\ell}| < 2.4$ and $p^{\ell}_T >
25$ GeV, to take into account the momentum carried away by the
neutrinos, impose a limit on the missing transverse energy $E^{\rm
miss}_{T}$, which is defined as a modulus of the vector sum of the
transverse momenta of the decay products, in inclusive reactions
$E^{\rm miss}_{T}$ is equal to transverse momentum of neutrino
$p^{\nu}_T$, in measurements $p^{\nu}_T > 20$ GeV. In addition, the
transverse energy of the $W^{\pm}$ boson $M^{W}_T$ must be greater
than 40 GeV, which is defined as $M^{W}_T = \sqrt{2 \, p_T^{\mu} \,
p_T^{\nu} \, (1 - \cos\Delta\phi_{\mu \nu})}$.

All data for associated $J/\psi$ plus $Z / W$ are presented as ratios of
associated production cross section $\sigma(J/\psi + Z / W)$ to
inclusive production cross section of the $Z / W^{\pm}$ boson
$\sigma(Z / W)$. In our calculations, we define the function
\begin{equation}\label{eq:Rfunction}
{\cal R}({\cal Q} + A) =
\frac{{\cal B({\cal Q} \to \mu \bar\mu)}}{\sigma(A + X)} \times
\frac{d \sigma({\cal Q} + A)}{d p^{\cal Q}_T},
\end{equation}
where ${\cal Q} = J/\psi$ or $\Upsilon(1S)$ and $A = Z$ or $W$,
inclusive cross sections $\sigma(Z / W + X)$ are from
Table.~\ref{table:1}, the values of two-muon branchings
${\cal B}(J/\psi \to \mu \bar\mu) = 0.0596$ and
${\cal B}(\Upsilon \to \mu \bar\mu) = 0.0248$ are used.

In the top panel of the Fig.~\ref{fig:1}, we plot the the ${\cal R}$
function (${\cal Q} = J/\psi, A = Z$), which demonstrates
$p_T^{\psi}$-dependence for the cross section of $J/\psi$ plus $Z$
boson associated production. The contributions from the SPS and the
DPS mechanisms are shown separately. Here and below, the grey boxes
around the central lines in the figures indicate upper and lower
limits of the cross section obtained due to the variation of the
hard scale $\mu$ by the factors $\xi = 2$ or $\xi = 1 / 2$ around
the central value of the hard scale $\mu =
\left(m_T^Z+m_T^{\psi}\right)/2, \ {\rm where} \ m_{T}^{Z} =
\sqrt{m_Z^2 + (p_{T}^{Z})^2}, \ {\rm and} \ m_{T}^{\psi} =
\sqrt{m_{\psi}^2 + (p_{T}^{\psi})^2}$. As it is estimated, the SPS
contribution dominates only at the very large $p_T^{\psi}$. The
ratio of the total cross section $\sigma^{\rm SPS} / \sigma^{\rm
DPS} \simeq 1 / 4$ at the $\sigma_{\rm eff} = 11$ mb. The
hadronization parameter ${\cal F}^{\psi} = 0.009$ taken
under similar kinematic conditions is used, i.e.
in the region of large $p_T^{\psi}$ and central rapidity interval.
The agreement between our prediction and experimental data looks
well excluding only difference at the large $p_T^{\psi}$.

In the bottom panel of the Fig.~\ref{fig:1}, we plot the normalized
$\Delta \phi_{\psi Z}$ spectrum. The spectrum is rather flat due to
the large DPS contribution, but at the $\Delta \phi_{\psi Z}\simeq \pi$,
it has peak in the data which is described well due to the SPS
contribution concentrated in this region of the $\Delta \phi_{\psi Z}$.

Our predictions for $J/\psi$ plus $Z$ associated production spectra
at the energy $\sqrt{s} = 13$ TeV are shown in the Fig.~\ref{fig:2}.
Parameter ${\cal F}^{\psi}$ is the same as at $\sqrt{s} = 8$ TeV,
$K$ factor from Table~\ref{table:1}.

In the top panel of the Figs.~\ref{fig:3} and \ref{fig:4}, we plot
function~(\ref{eq:Rfunction}) (${\cal Q} = J/\psi, \; A = W$) in
comparing with the experimental data~\cite{ATLAS:PsiW7,ATLAS:PsiW8}.
Taking into consideration errors of the theoretical calculations
coming from the variation of the hard scale $\mu$, we conclude that
the agreement between data and our theoretical calculations in the
ICEM  and the PRA, is even more precise than in case of associated
$J/\psi$ plus $Z$ boson production.  In the Fig.~\ref{fig:5}, we
present the predictions for cross section of $J/\psi$ plus $W$
associated production at $\sqrt{s} = 13$ TeV.

Recently, we have performed calculations in the ICEM using PRA for
the single and double prompt $\Upsilon=\Upsilon(1S)$ meson
production at the LHC~\cite{ChernyshevSaleev2Ups2022}. We found good
description of the data taking into account the SPS and the DPS
production scenarios, with the universal value of DPS parameter
$\sigma_{\rm eff} = 11$ mb. Predictions of the prompt $\Upsilon$
plus $Z (W)$ associated production spectra as a function of
$p_T^{\Upsilon}$ and $\Delta \phi_{\Upsilon Z}$ at $\sqrt{s} = 8
\mbox{ and } 13$ TeV are presented in the Figs.~\ref{fig:6}.
Kinematical cuts on $\Upsilon$ mesons are the same as for $J/\psi$
mesons from Table~\ref{table:1}. The hadronization parameter ${\cal
F}^{\Upsilon} = 0.021$ is taken from the fit of single $\Upsilon$
production~\cite{ChernyshevSaleev2Ups2022}.

\section{Conclusions}
\label{sec:conclusions}

We obtain a quite satisfactory description for the prompt $J/\psi$
mesons $p_T$ spectra in the associated $J/\psi$ plus $Z ( W)$ bosons
production in the ICEM using the PRA at the energy of the LHC, as it
was measured by the ATLAS
Collaboration~\cite{ATLAS:PsiZ,ATLAS:PsiW7,ATLAS:PsiW8}. The
azimuthal angle difference spectra as functions of $\Delta
\phi_{\psi Z}$ and $\Delta \phi_{\psi W}$ are well described too.
Both mechanisms, SPS and DPS, have been considered. We don't use any
free parameters to obtain our prediction, the relevant ones,
$\sigma_{\rm eff}$, ${\cal F}^{\psi}$ and ${\cal F}^{\Upsilon}$,
have been fixed early, when we described single and double prompt
$J/\psi$ and $\Upsilon$ production at the
LHC~\cite{ChernyshevSaleev2Psi2022,ChernyshevSaleev2Ups2022}. We
find the dominant role of the DPS production mechanism in the
considered here processes of the associated production, such as
$\sigma^{\rm SPS} / \sigma^{\rm DPS} = 0.2\sim 0.25$. The
predictions for the cross sections and spectra at the energies
$\sqrt{s} = 8, 13$ TeV have been done.

\section*{Acknowledgments}
We are grateful to A.~Van~Hameren for advice on the program KaTie
and M.~Nefedov for the helpful communication.

\bibliographystyle{ws-ijmpa}
\bibliography{references_Z}

\pagebreak

\begin{figure}[ht]
\includegraphics[scale=0.3]{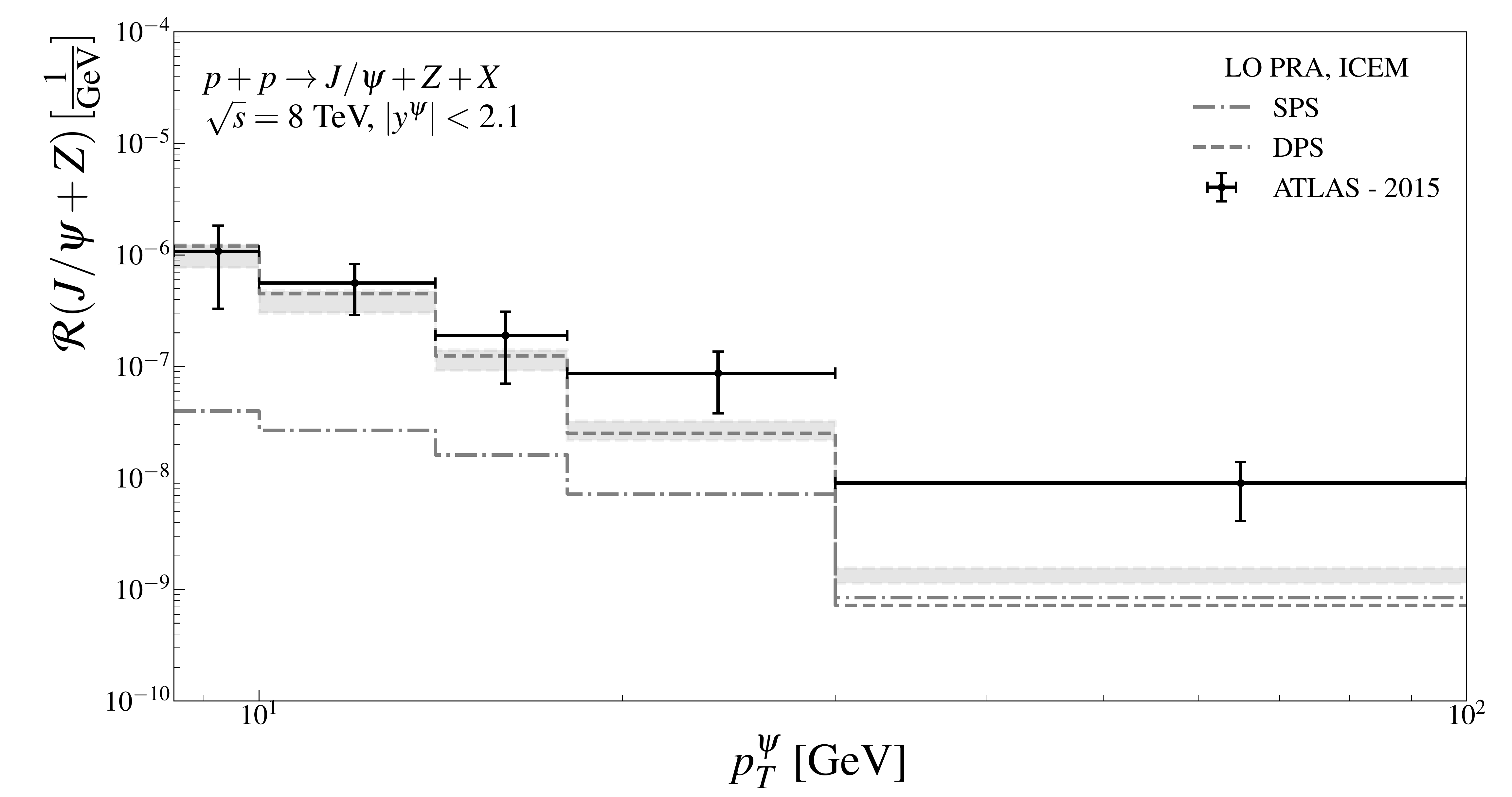}
\includegraphics[scale=0.3]{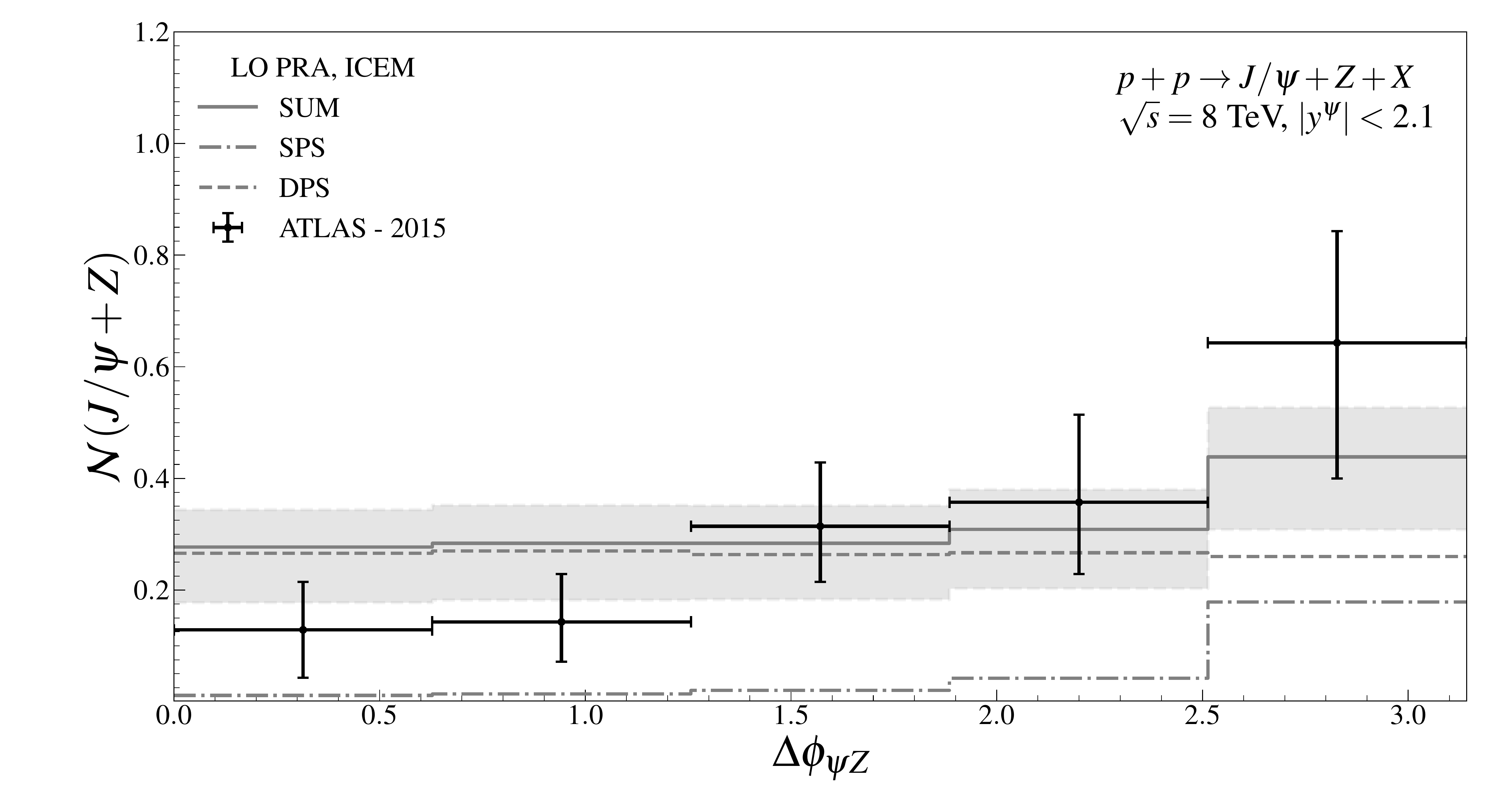}

 \caption{Spectra of $J/\psi$ plus $Z$ boson associated production
as a function of $J/\psi$ transverse momenta $p_T^{\psi}$ (top
panel) and the azimuthal angle difference $\Delta \phi_{\psi Z}$
(bottom panel) at $\sqrt{s} = 8$ TeV. The data are from the ATLAS
Collaboration~\cite{ATLAS:PsiZ}.} \label{fig:1}
\end{figure}

\begin{figure}[ht]
\includegraphics[scale=0.3]{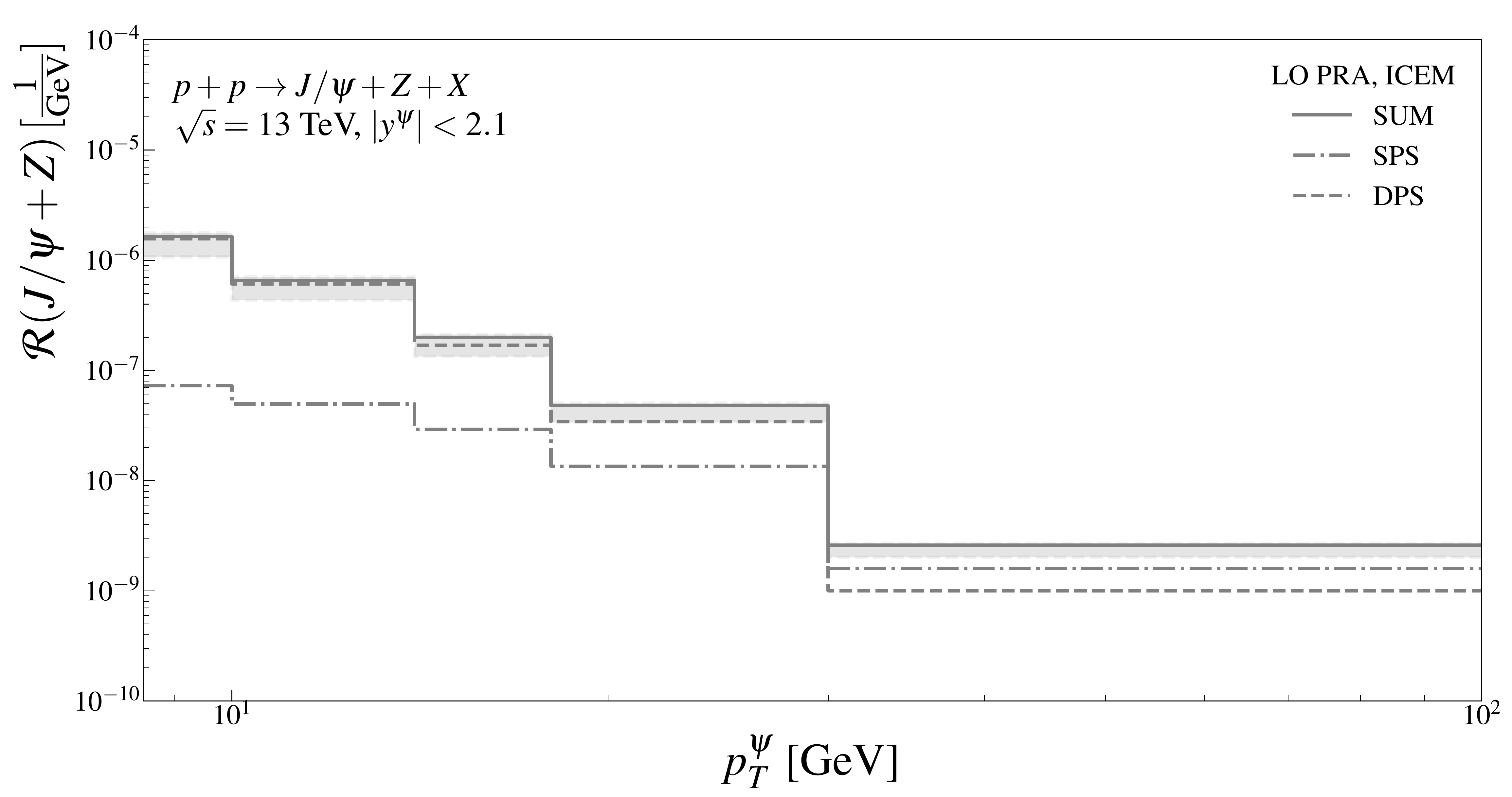}
\includegraphics[scale=0.3]{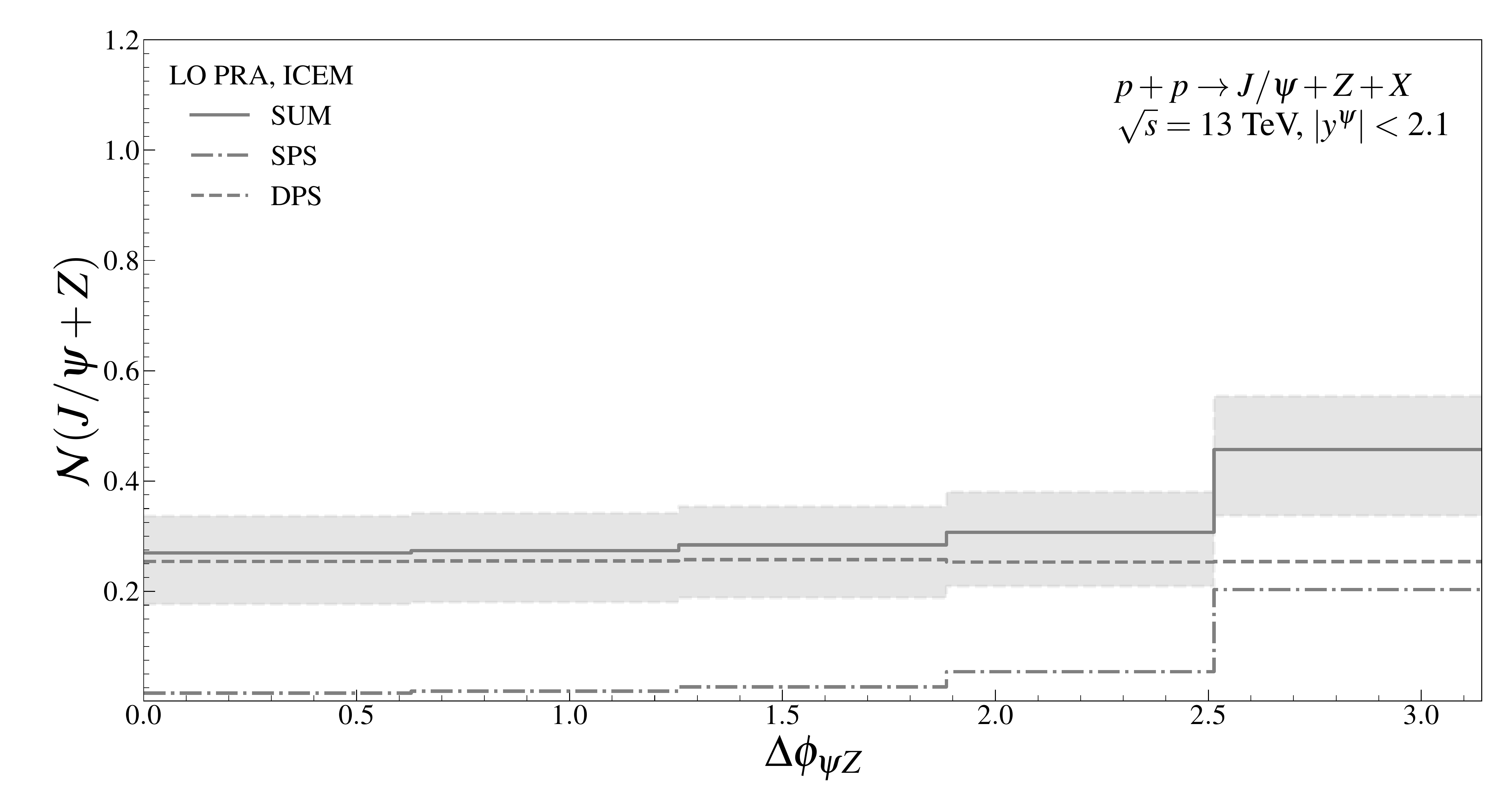}

\caption{Predictions for $J/\psi$ plus $Z$ boson associated
production spectra as a function of $J/\psi$ transverse momenta
$p_T^{\psi}$ (top panel) and the azimuthal angle difference $\Delta
\phi_{\psi Z}$ (bottom panel) at $\sqrt{s} = 13$ TeV.} \label{fig:2}
\end{figure}

\begin{figure}[ht]

\includegraphics[scale=0.3]{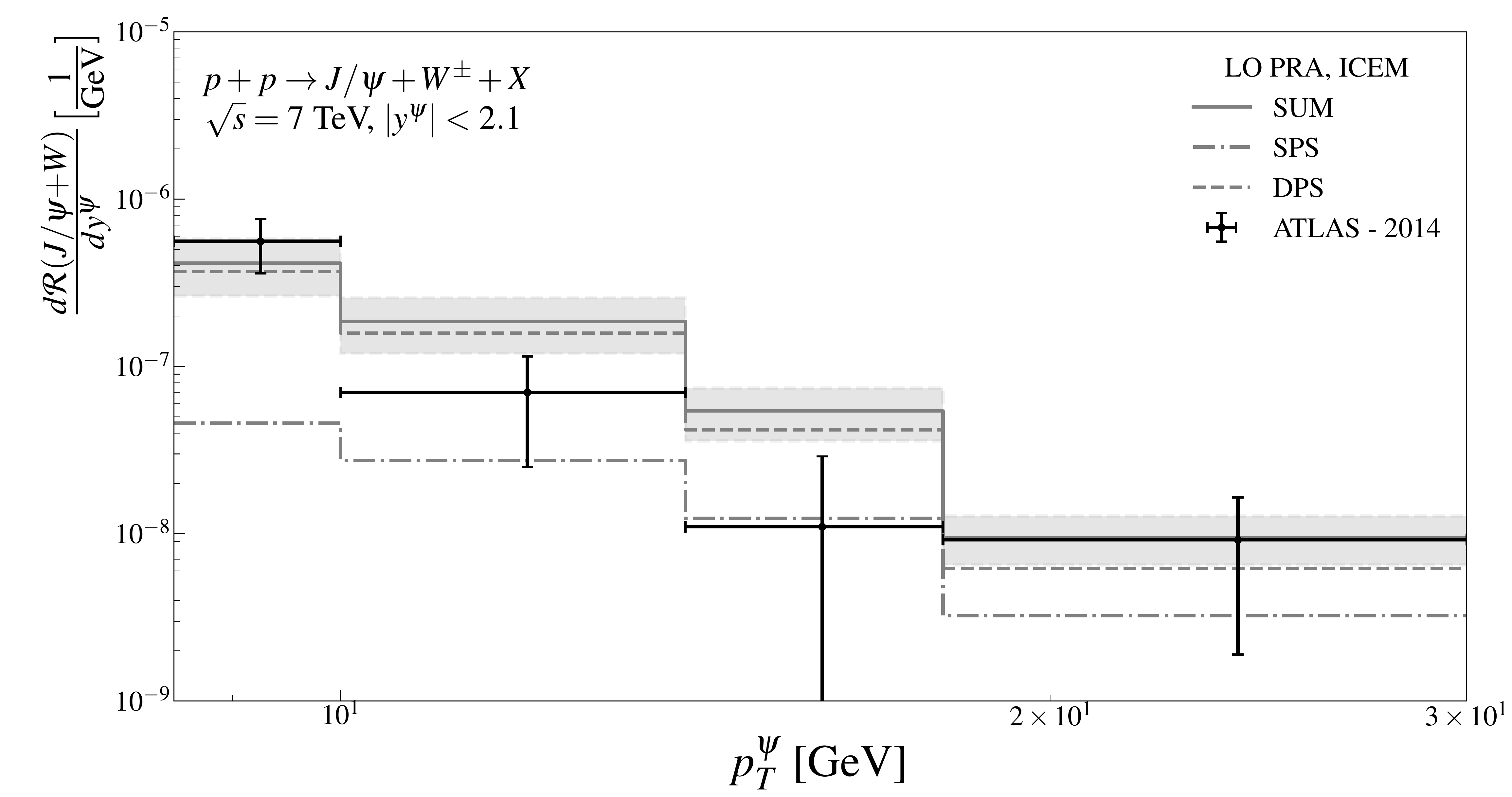}
\includegraphics[scale=0.3]{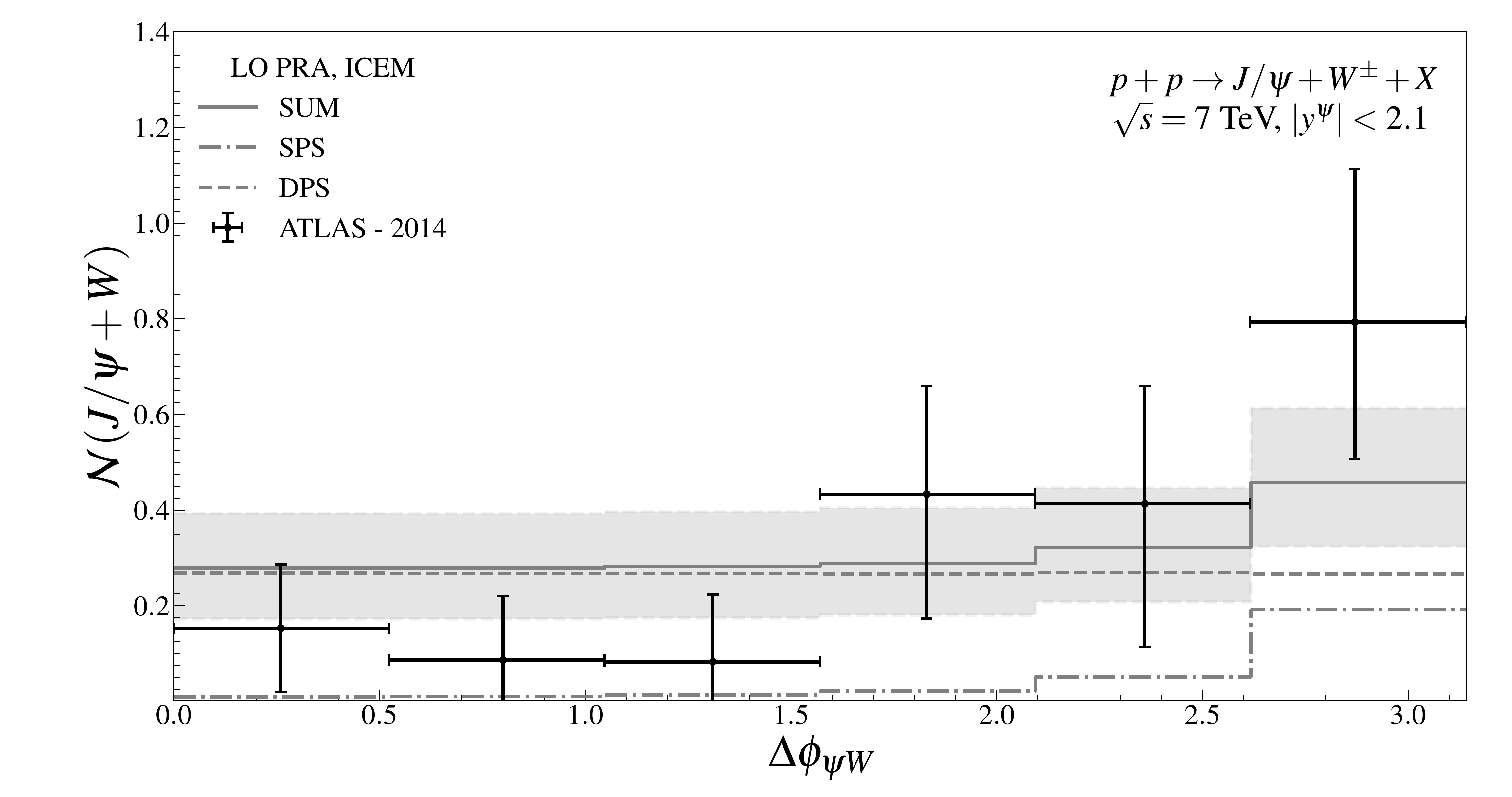}

\caption{Spectra of $J/\psi$ plus $W^{\pm}$ associated production as a
function of $p_T^{\psi}$ (top panel) and $\Delta \phi_{\psi W}$
(bottom panel) at $\sqrt{s} = 7$ TeV. The data are from the ATLAS
Collaboration~\cite{ATLAS:PsiW7}.} \label{fig:3}

\end{figure}

\begin{figure}[ht]

\includegraphics[scale=0.3]{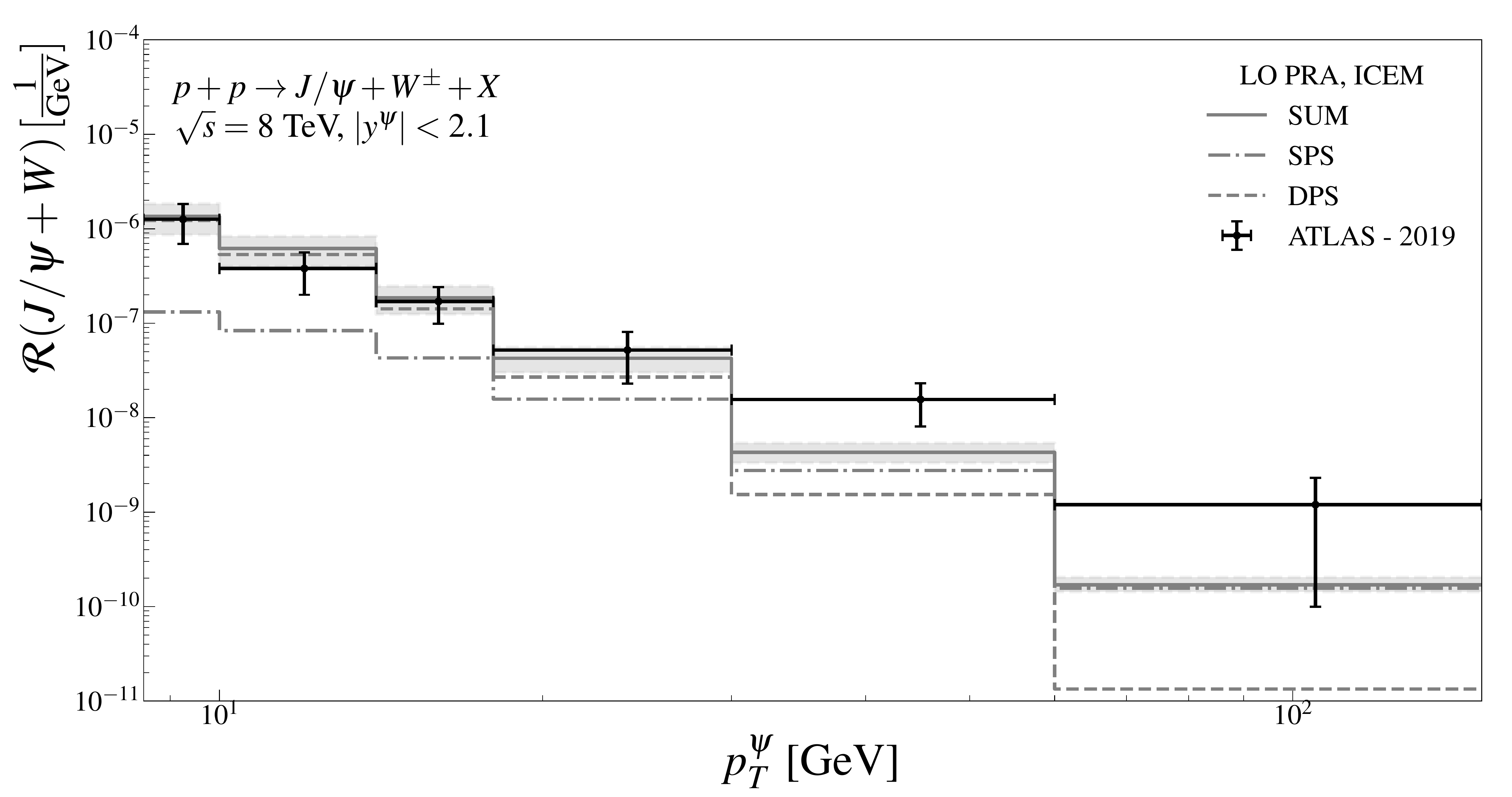}
\includegraphics[scale=0.3]{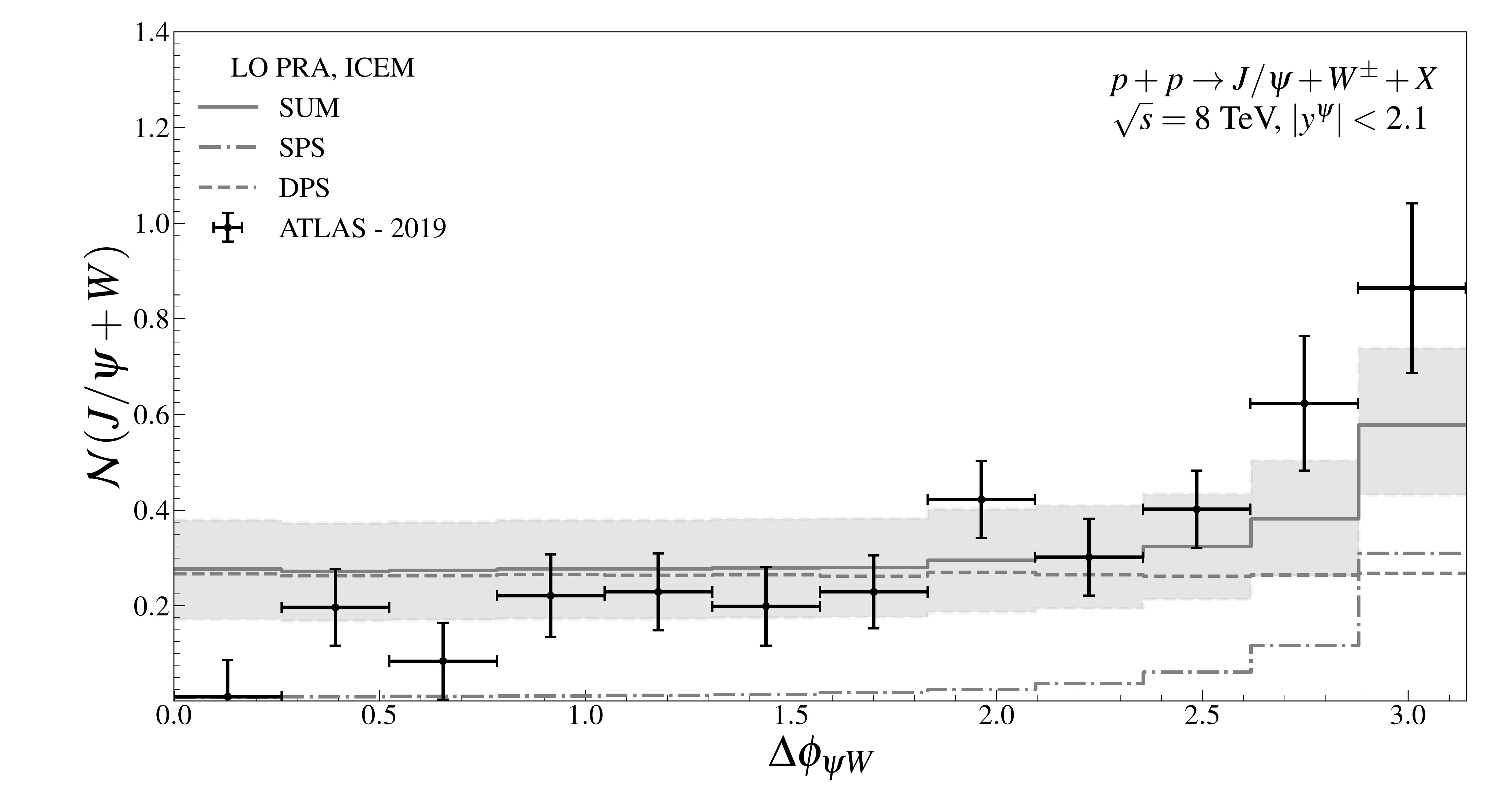}

 \caption{Spectra of $J/\psi$ plus $W^{\pm}$ associated production as a
function of $p_T^{\psi}$ (top panel) and $\Delta \phi_{\psi W}$
(bottom panel) at $\sqrt{s} = 8$ TeV. The data are from the ATLAS
Collaboration~\cite{ATLAS:PsiW8}.} \label{fig:4}

\end{figure}

\begin{figure}[ht]

\includegraphics[scale=0.3]{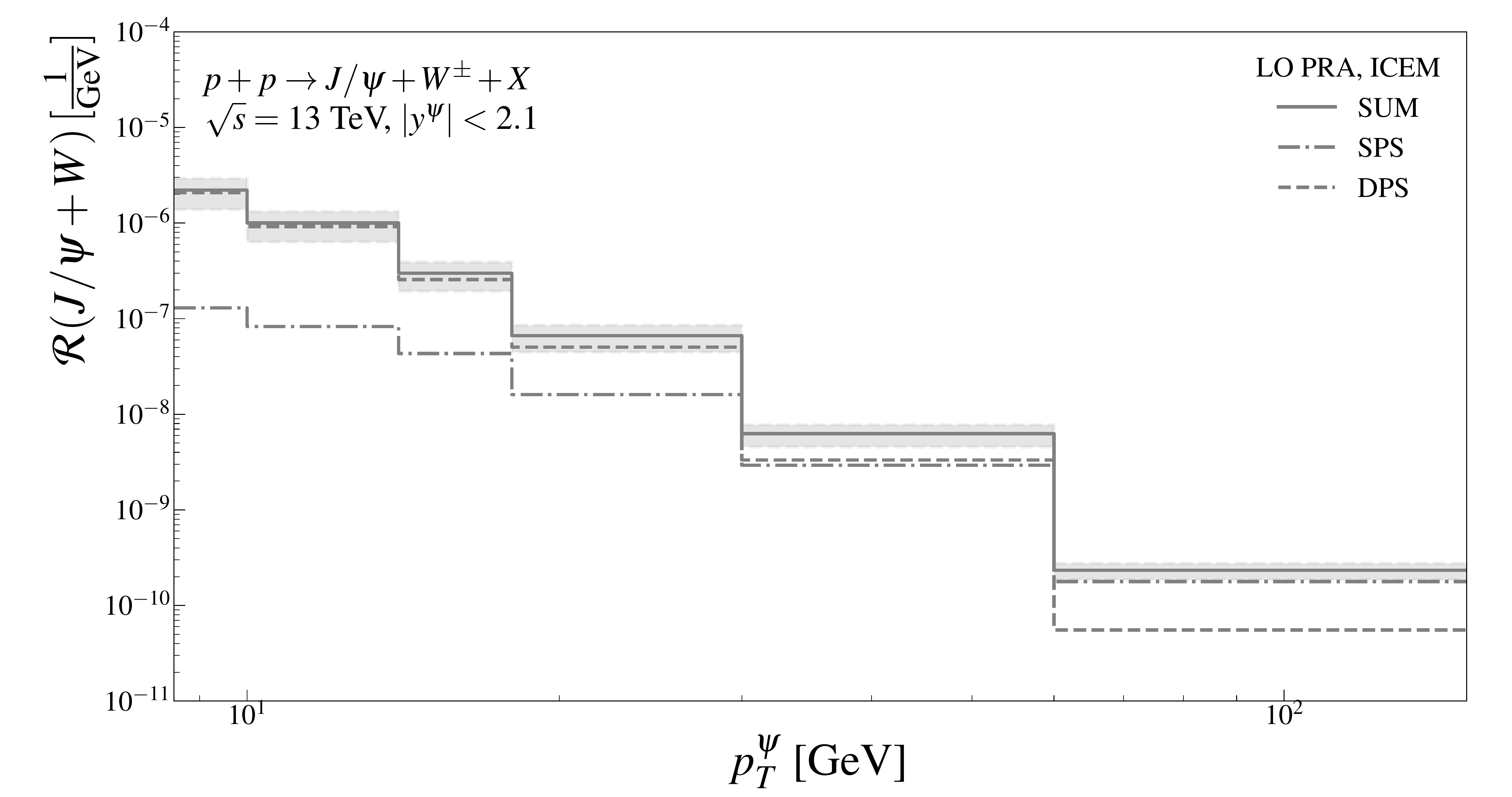}
\includegraphics[scale=0.3]{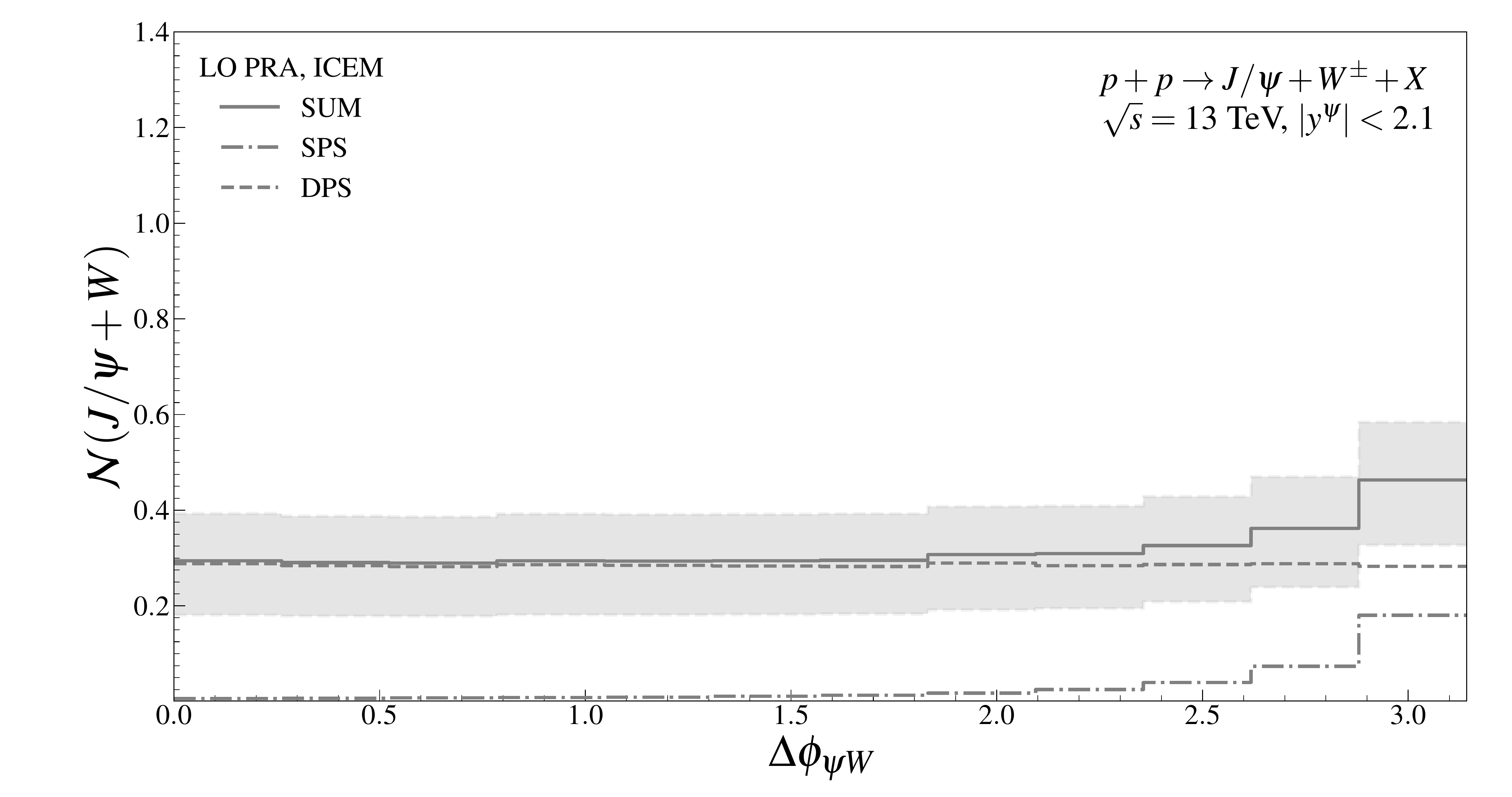}
 \caption{Predictions for $J/\psi$ plus $W^{\pm}$ associated
production spectra as a function of $p_T^{\psi}$ and
$\Delta \phi_{\psi W}$ at $\sqrt{s} = 13$ TeV.} \label{fig:5}

\end{figure}

\begin{figure}[ht]

\includegraphics[scale=0.3]{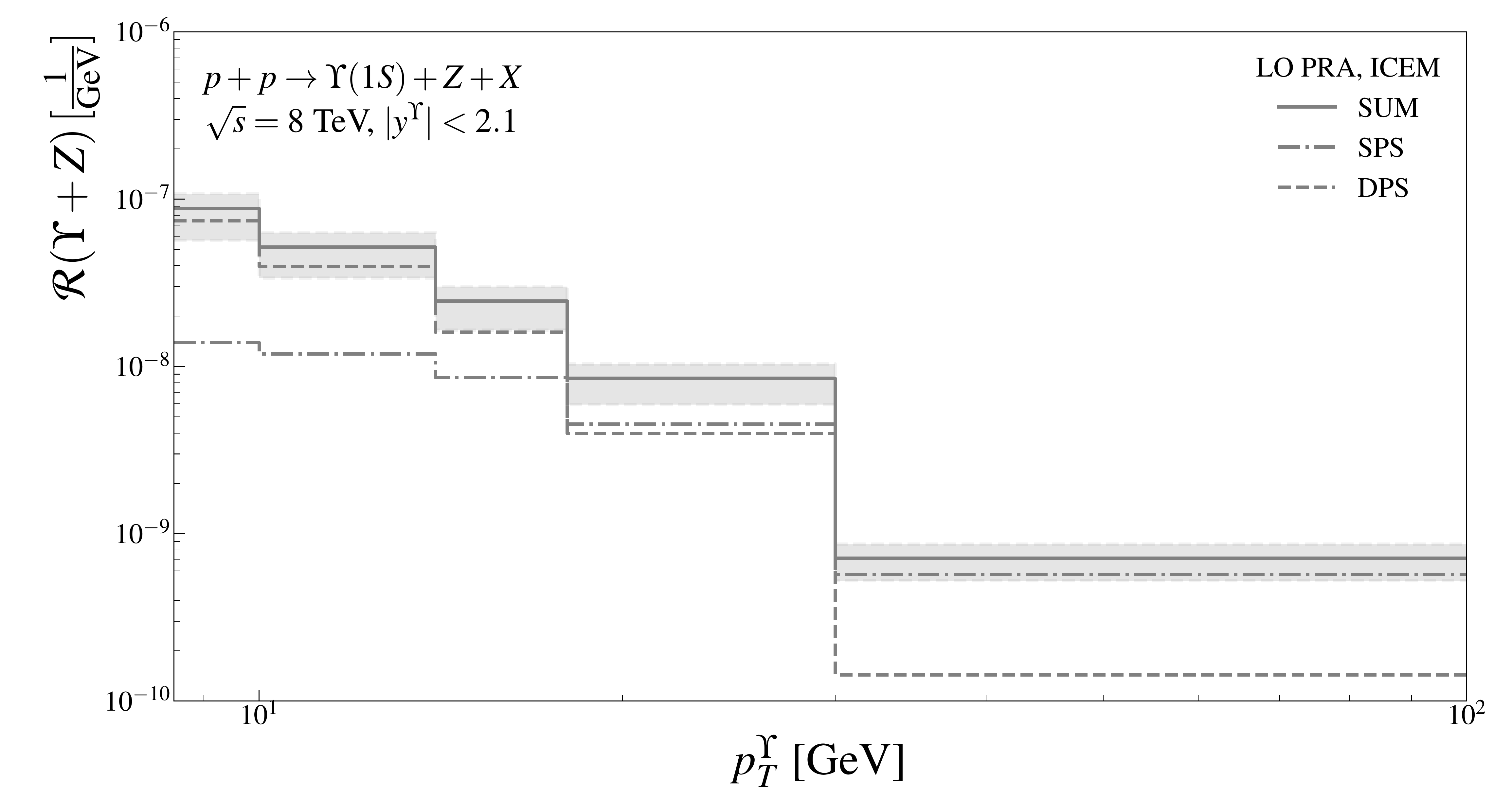}
\includegraphics[scale=0.3]{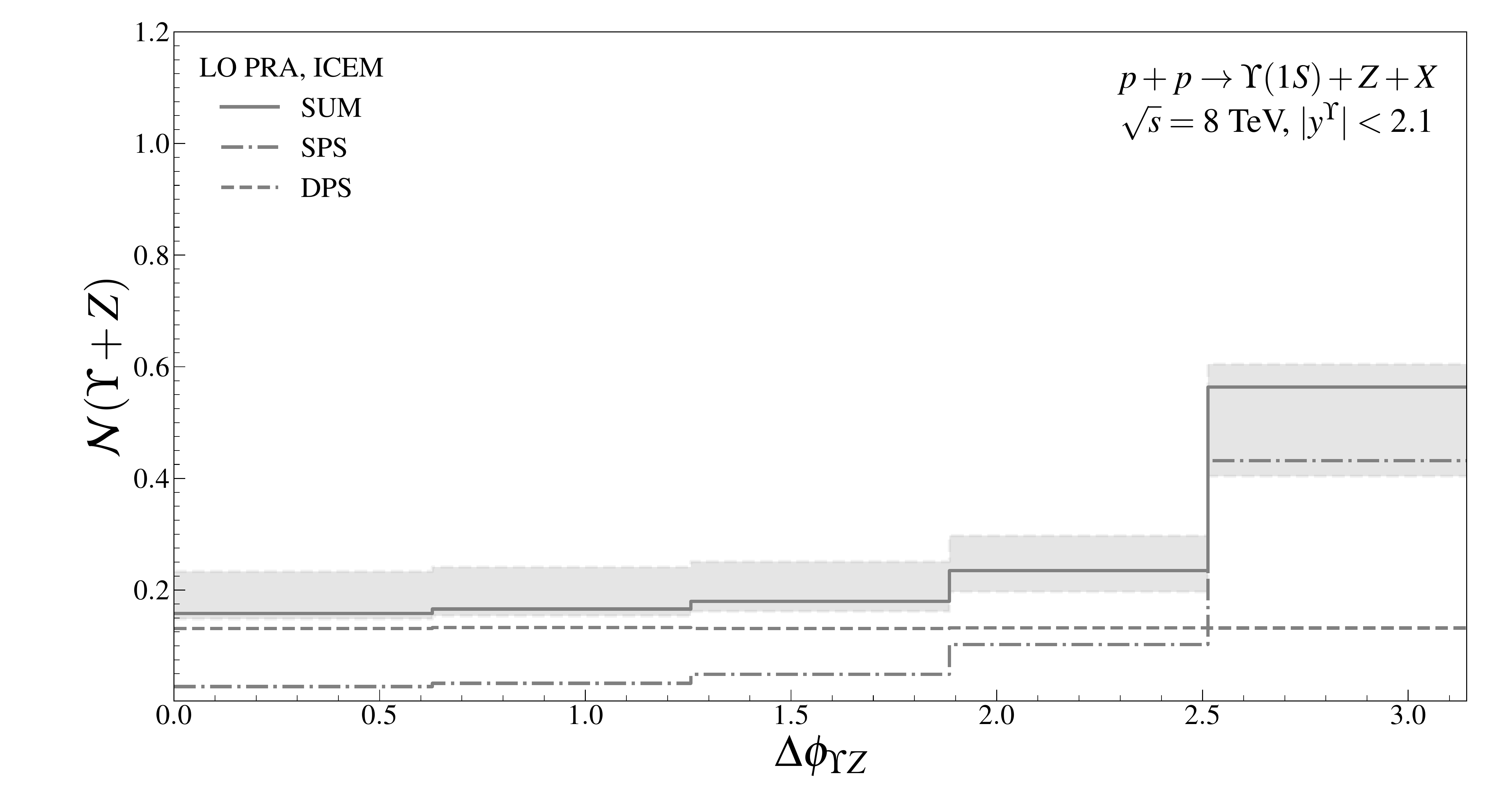}

 \caption{Spectra of $\Upsilon$ plus $Z$ boson associated
production as a function of $\Upsilon$ transverse momenta
$p_T^{\Upsilon}$ (top panel) and the azimuthal angle difference
$\Delta \phi_{\Upsilon Z}$ (bottom panel) at $\sqrt{s} = 8$ TeV.}
\label{fig:6}

\end{figure}

\begin{figure}[ht]

\includegraphics[scale=0.3]{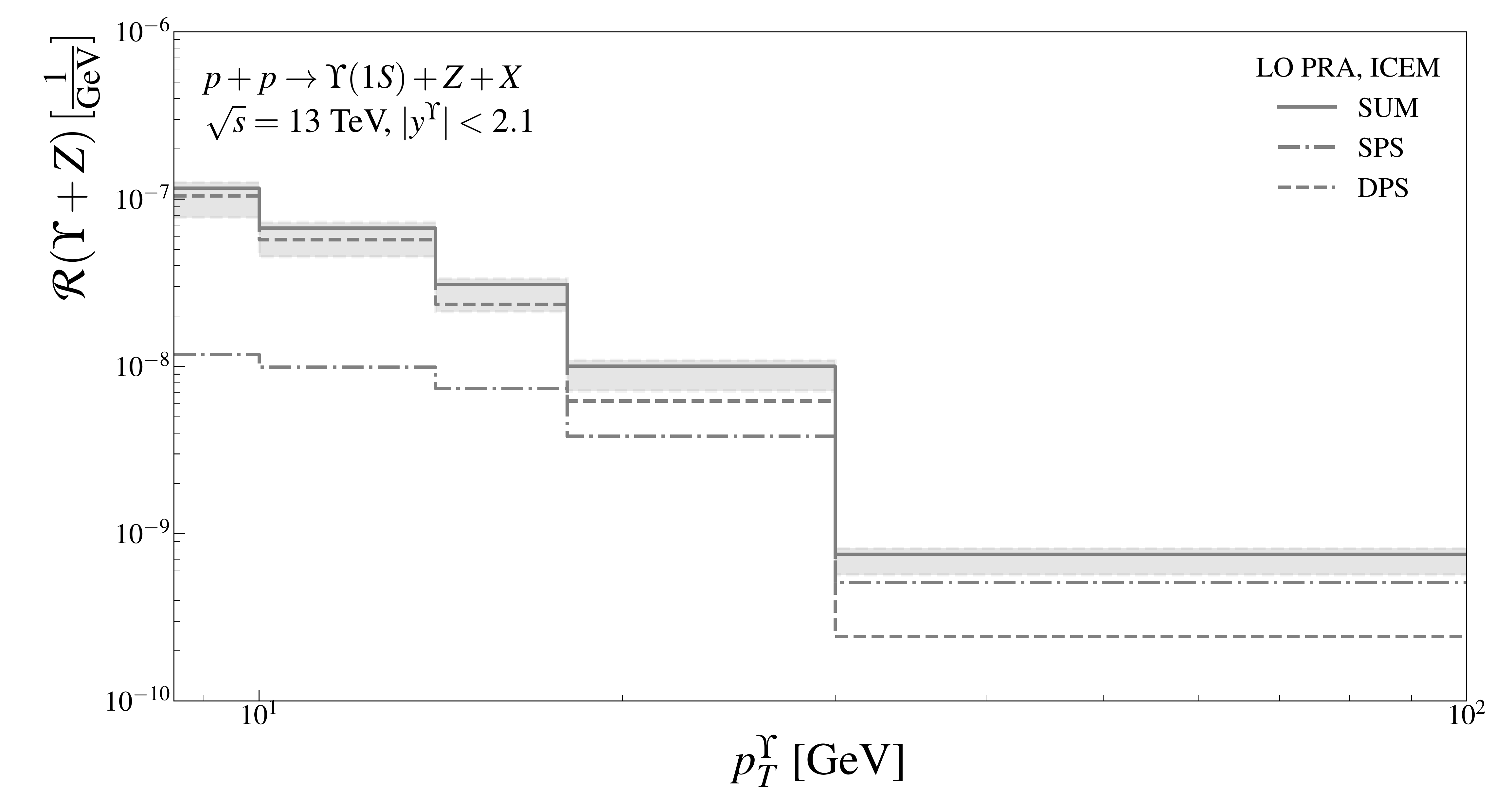}
\includegraphics[scale=0.3]{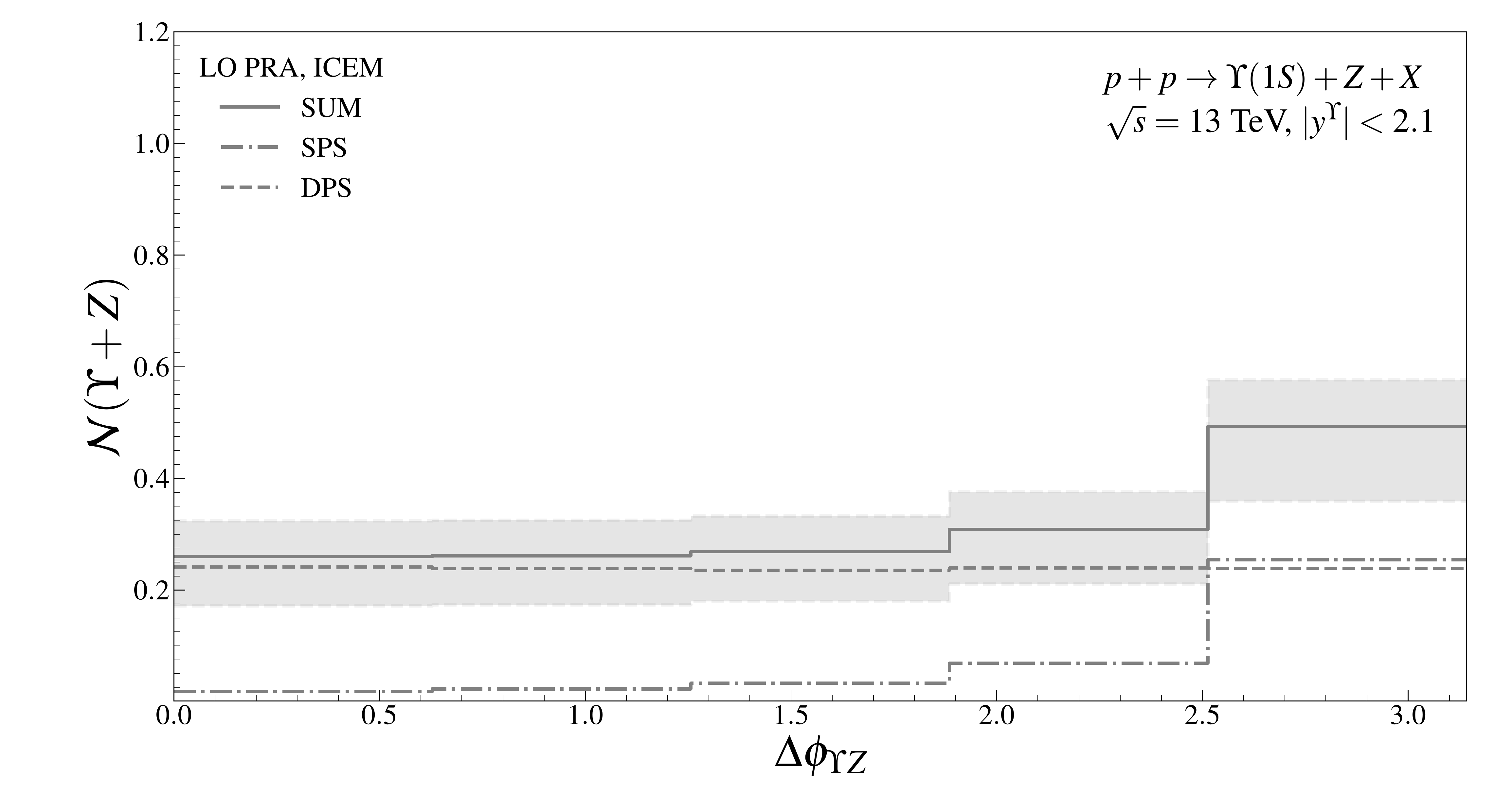}

 \caption{Spectra of $\Upsilon$ plus $Z$ boson associated
production as a function of $\Upsilon$ transverse momenta
$p_T^{\Upsilon}$ (top panel) and the azimuthal angle difference
$\Delta \phi_{\Upsilon Z}$ (bottom panel) at $\sqrt{s} = 13$ TeV.}
\label{fig:7}

\end{figure}

\begin{figure}[ht]

\includegraphics[scale=0.3]{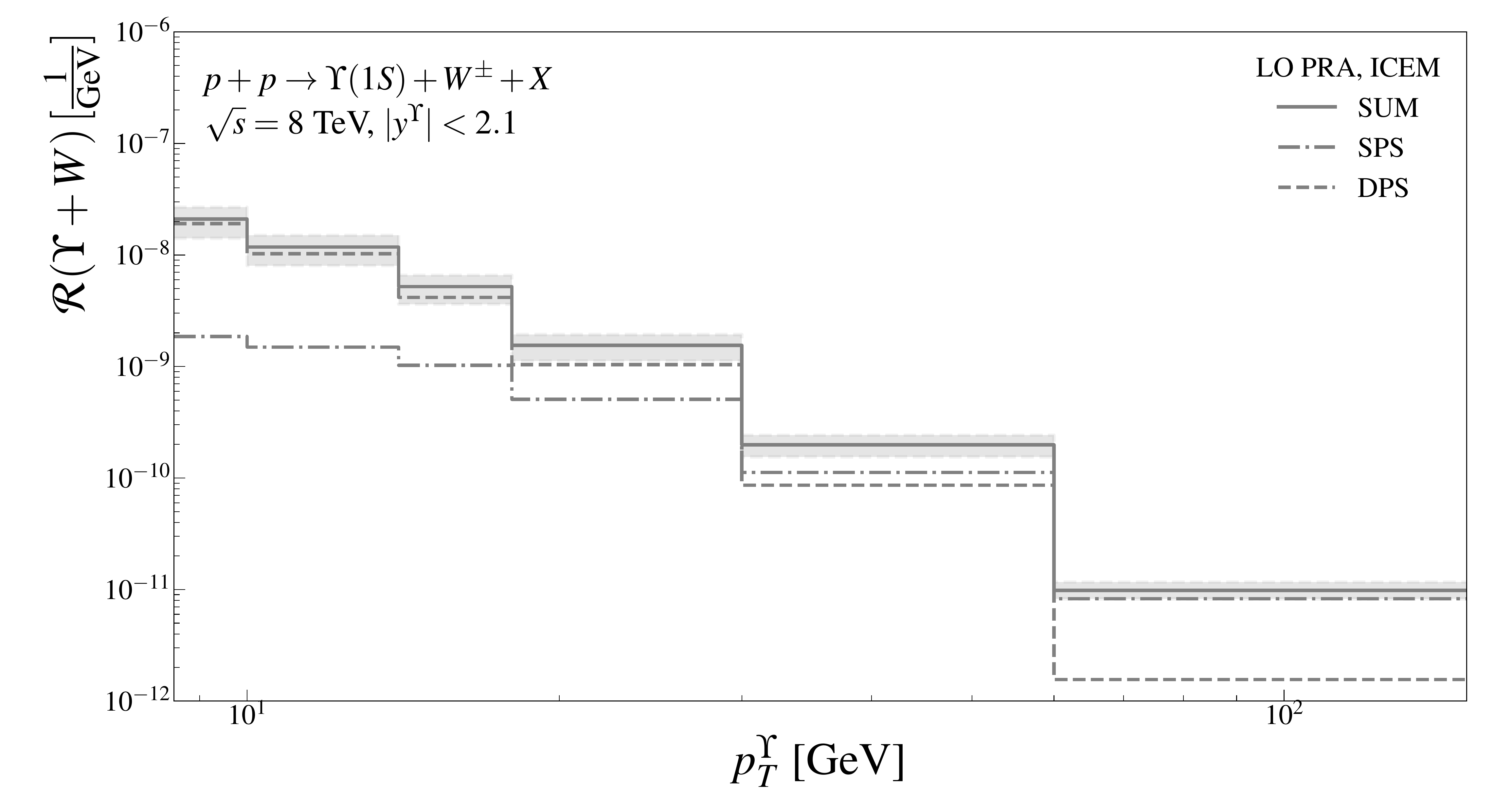}
\includegraphics[scale=0.3]{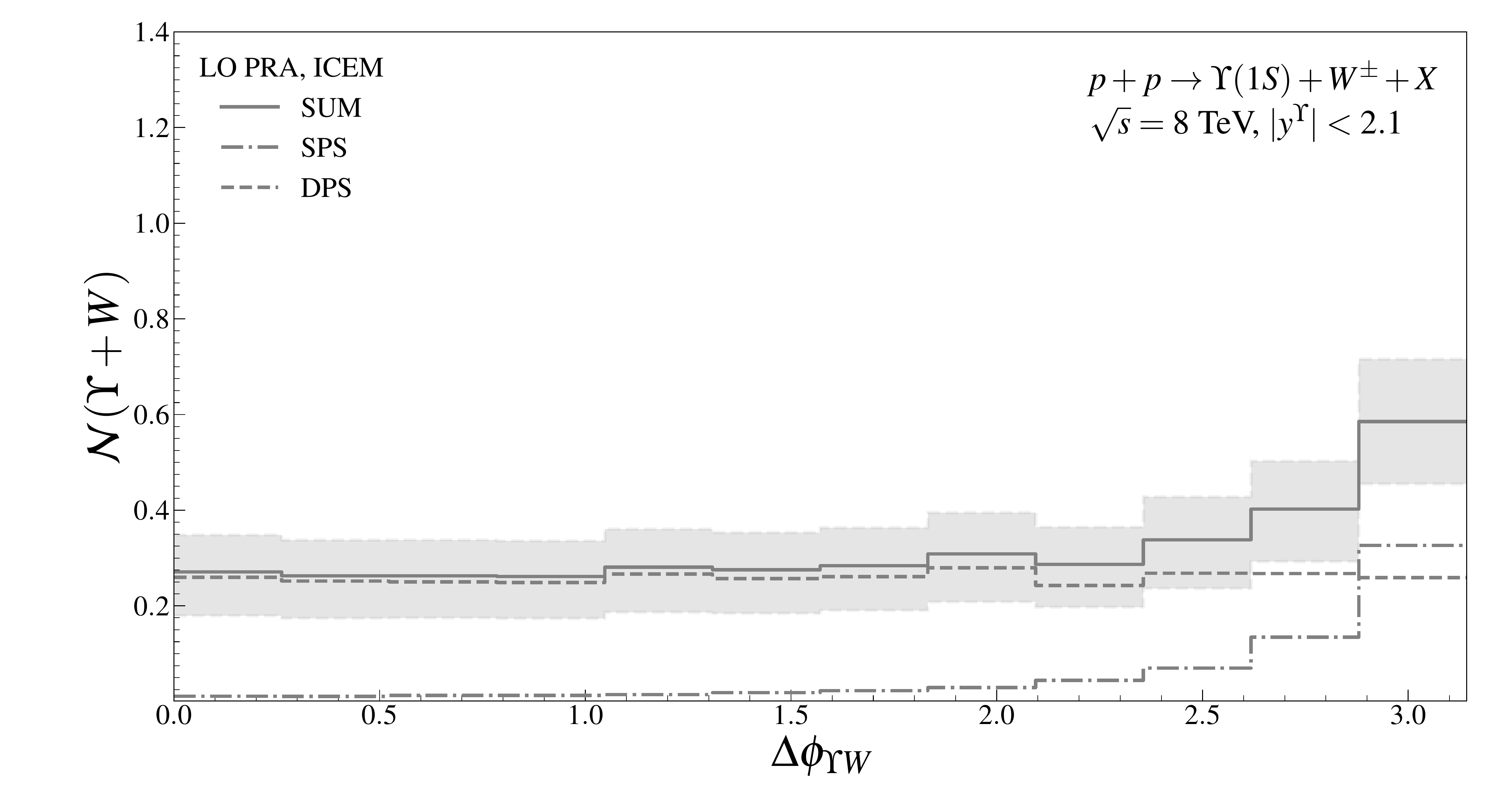}

 \caption{Spectra of $\Upsilon$ plus $W^{\pm}$ boson associated
production as a function of $\Upsilon$ transverse momenta
$p_T^{\Upsilon}$ (top panel) and the azimuthal angle difference
$\Delta \phi_{\Upsilon W}$ (bottom panel) at $\sqrt{s} = 8$ TeV.}
\label{fig:8}

\end{figure}

\begin{figure}[ht]

\includegraphics[scale=0.3]{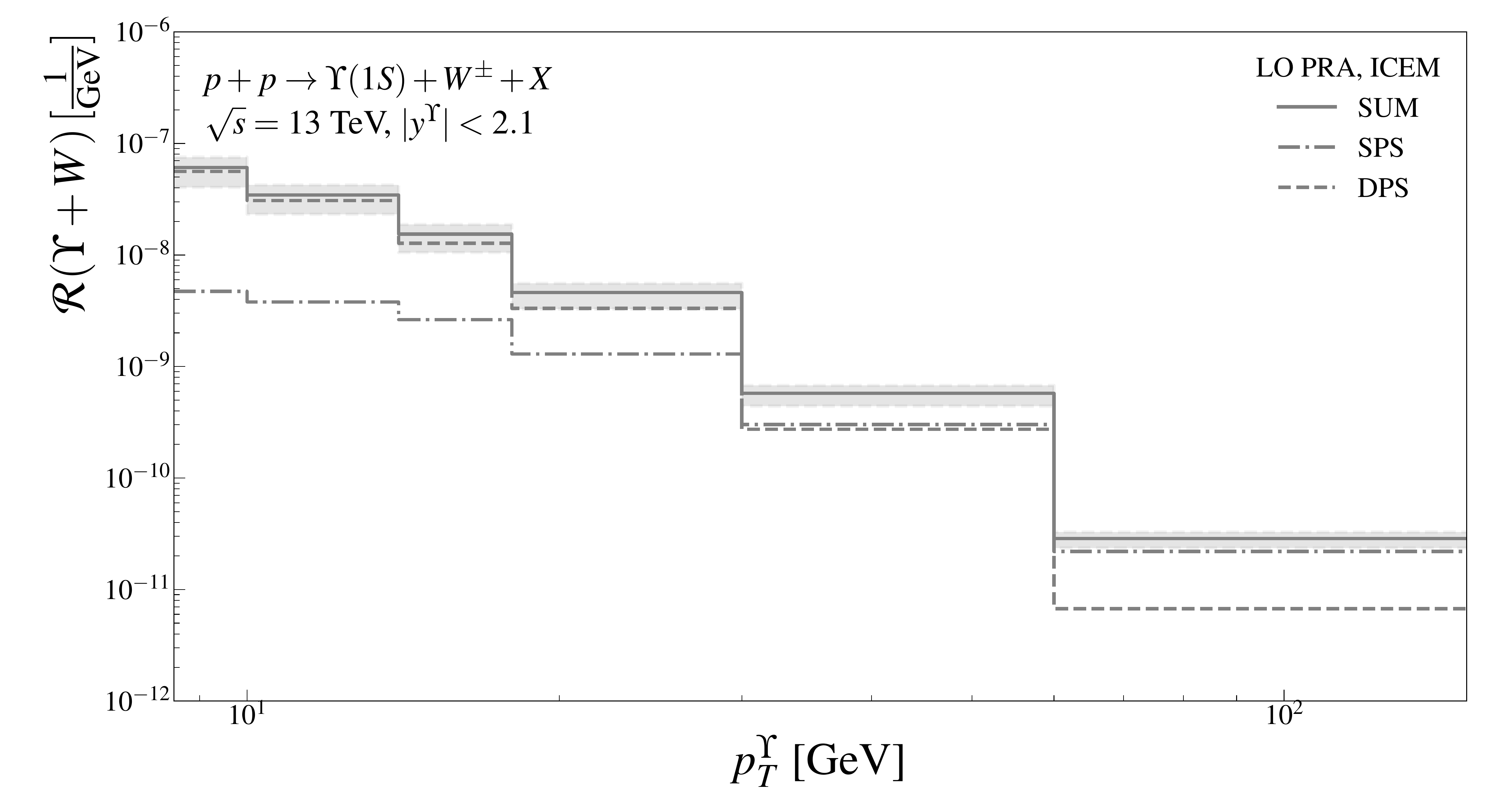}
\includegraphics[scale=0.3]{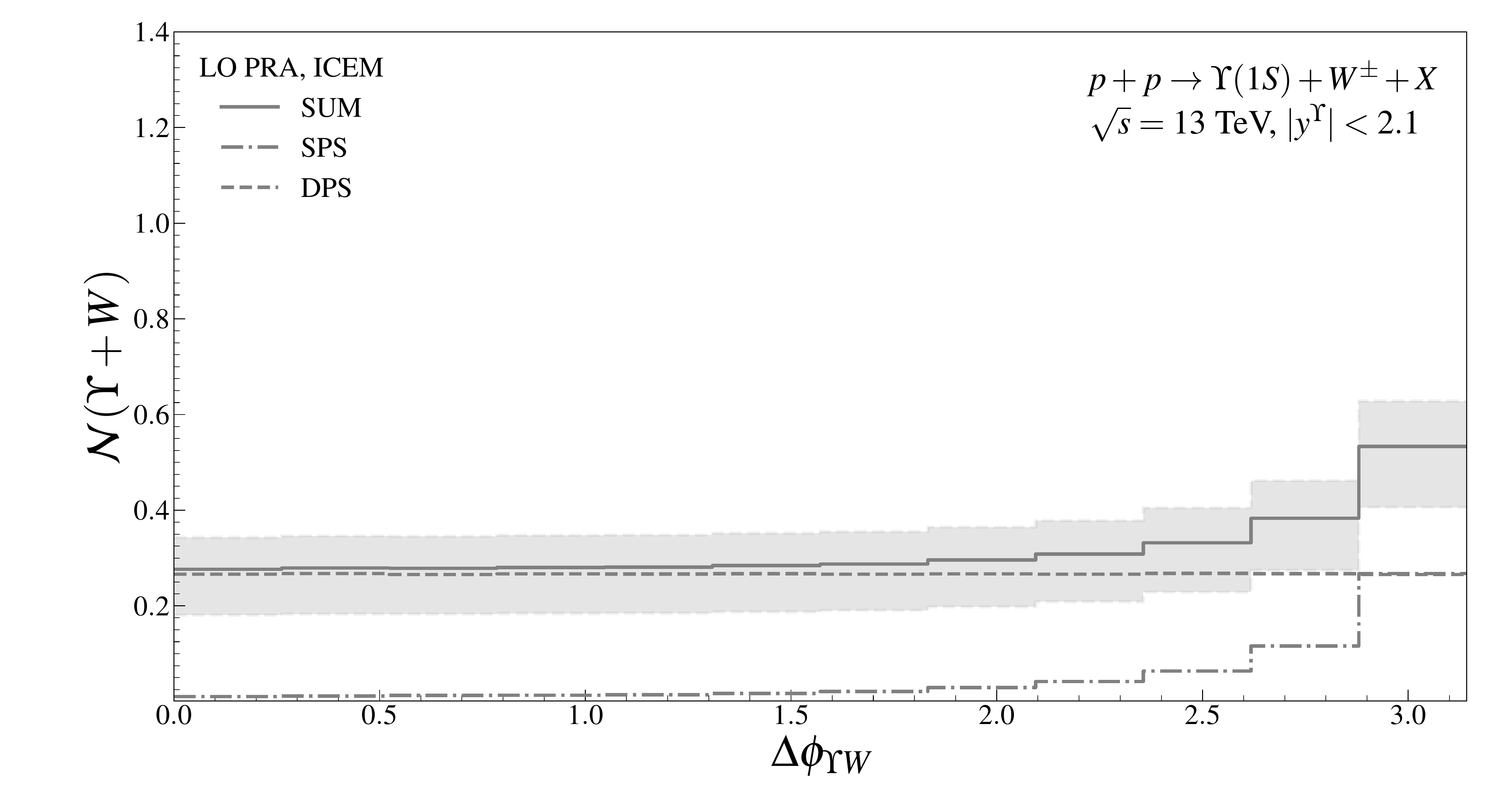}

 \caption{Spectra of $\Upsilon$ plus $W^{\pm}$ boson associated
production as a function of $\Upsilon$ transverse momenta
$p_T^{\Upsilon}$ (top panel) and the azimuthal angle difference
$\Delta \phi_{\Upsilon W}$ (bottom panel) at $\sqrt{s} = 13$ TeV.}
\label{fig:9}

\end{figure}

\end{document}